\documentclass{article}
\usepackage{spconf,amsmath,graphicx}
\usepackage{cite}
\usepackage{booktabs}

\title{HASA-Net: A NON-INTRUSIVE HEARING-AID SPEECH ASSESSMENT NETWORK }
\name{Hsin-Tien Chiang$^{1}$, Yi-Chiao Wu$^{2}$, Cheng Yu$^{1}$, Tomoki Toda$^{2}$, Hsin-Min Wang$^{3}$, Yih-Chun Hu$^{4}$, Yu Tsao$^{1}$}
\address{$^{1}$Research Center for Information Technology Innovation, Academia Sinica, Taiwan\\
$^{2}$Information Technology Center, Nagoya University, Japan\\
$^{3}$Institute of Information Science, Academia Sinica, Taiwan\\
$^{4}$University of Illinois at Urbana-Champaign, USA}
\begin{document}
%
\maketitle
\begin{abstract}
Without the need of a clean reference, non-intrusive speech assessment methods have caught great attention for objective evaluations. Recently, deep neural network (DNN) models have been applied to build non-intrusive speech assessment approaches and confirmed to provide promising performance. However, most DNN-based approaches are designed for normal-hearing listeners without considering hearing-loss factors. In this study, we propose a DNN-based hearing aid speech assessment network (HASA-Net), formed by a bidirectional long short-term memory (BLSTM) model, to predict speech quality and intelligibility scores simultaneously according to input speech signals and specified hearing-loss patterns. To the best of our knowledge, HASA-Net is the first work to incorporate quality and intelligibility assessments utilizing a unified DNN-based non-intrusive model for hearing aids. Experimental results show that the predicted speech quality and intelligibility scores of HASA-Net are highly correlated to two well-known intrusive hearing-aid evaluation metrics, hearing aid speech quality index (HASQI) and hearing aid speech perception index (HASPI), respectively.
\end{abstract}
\section{Introduction}
\label{sec:intro}
Speech quality and intelligibility assessments serve as important tools to many speech-related applications. Speech quality indicates pleasantness or naturalness of a speech signal, and speech intelligibility measures how well the content of the speech can be understood. A straightforward approach to measure speech quality or intelligibility is to conduct subjective listening tests. In such test, speech signals are played to a group of listeners, who are then asked to score the quality or provide recognized words of the heard speech signals. To avoid potential assessment biases, a large number of subjects are generally required. However, conducting listening tests on many subjects is time-consuming and prohibitively expensive. Therefore, objective speech quality and intelligibility assessments are developed and used as surrogates to the subjective listening tests.

Objective speech assessments can be roughly divided into two categories, namely, intrusive and non-intrusive. Intrusive methods \cite{rix2001perceptual,vincent2006performance,taal2011algorithm,kates2014hearing1,kates2014hearing2,le2019sdr} use the clean speech as reference to compare with the degraded/processed one to give the output index. However, clean speech may not always be available and thus restrict their practicality \cite{islam2016non}. On the other hand, non-intrusive methods \cite{malfait2006p,kim2007anique+,falk2010non} are capable of calculating the output index directly on the degraded/processed speech without reference data. Without the requirement of reference data, non-intrusive methods have been widely adopted in online assessment and real-world applications.

Recently, DNN models have been used as a fundamental tool for speech quality and intelligibility assessment \cite{fu2018quality,lo2019mosnet,dong2020attention,zezario2020stoi,leng2021mbnet}. For speech quality, Quality-Net \cite{fu2018quality} was proposed as an end-to-end, non-intrusive speech quality evaluation model; Quality-Net is a BLSTM-based model and capable to predict perceptual evaluation of speech quality (PESQ) scores for noisy/processed speech signals. MOSNet \cite{lo2019mosnet} was designed to predict mean opinion score (MOS) for converted speech. Later on, MBNet \cite{leng2021mbnet}, which consists of a MeanNet and a BiasNet, was proposed as an improved version of MOSNet, considering score variations caused by personal preferences. For speech intelligibility, STOI-Net \cite{zezario2020stoi} utilizes CNN-BLSTM with a multiplicative attention mechanism to predict speech intelligibility. A unified model that estimates multiple objective speech quality and intelligibility scores is developed and presented in \cite{dong2020attention}.

Despite the recent attention on utilizing DNN for objective speech evaluation, limited research considers hearing impaired listeners; that is, users with assistive listening devices, such as hearing aids and cochlear implants. Hearing  loss is ranked as the fourth highest cause of disability globally \cite{cunningham2017hearing} and refers to the total or partial inability to hear. The impacts of hearing loss have profound effects on ability to communicate. Approximately 80 percent of people were living with varying degrees of hearing loss in 2018 \cite{olusanya2019hearing}. The prevalence of hearing loss in the United States doubles with every ten year increase in age \cite{cunningham2017hearing}. In order to broaden the impact of speech research to include the elderly and hearing-impaired, research with hearing impaired listeners are of increased importance. So far, numerous advanced signal processing methods have been tested on scenarios of hearing impaired listeners \cite{lai2016deep, wang2020improving, healy2019optimal, nossier2019enhanced}. An effective speech quality and intelligibility assessment system for varying hearing-loss patterns thus becomes an essential tool to further develop suitable speech signal processing methods for assistive listening devices.   

In this study, we propose a novel DNN-based end to-end, non-intrusive hearing aid speech assessment model, which we called HASA-Net. HASA-Net is a unified model capable to simultaneously estimating quality and intelligibility given input speech and specified hearing-loss patterns. Specifically, HASA-Net aims to predict HASQI \cite{kates2014hearing1} and HASPI\cite{kates2014hearing2} scores, which are two well-known evaluation metrics for speech quality and intelligibility designed for hearing-impaired listeners. HASQI and HASPI are based on an auditory model that incorporates changes due to hearing loss. While both HASQI and HASPI are based on a calculation comparing the noisy signal with its clean reference, HASA-Net predicts scores \emph{without the need of clean reference}. To the best of our knowledge, HASA-Net represents the first step towards an end-to-end, non-intrusive unified model for listeners with hearing aids that simultaneously measures quality and intelligibility.  

The remainder of this paper is organized as follows. We describe the details of HASA-Net in Section 2. The experimental setup and results are presented in Section 3. Finally, we conclude this work in Section 4.

\section{HASA-Net}
\label{sec:relatedwork}
HASA-Net is an objective speech assessment model adopting multi-task learning and multi-head attention mechanism. We first introduce multi-task learning, multi-head attention mechanism, the overall framework of HASA-Net, and the loss function.

\subsection{Multi-task learning}
\label{ssec:mtk}
Multi-task learning (MTL) \cite{caruana1997multitask} has been widely adopted in speech-realted applications \cite{wu2015deep,chen2015speech,kim2017joint}. The aim of MTL is to simultaneously learn multiple related tasks with a unified model to obtain more robust shared feature representations, thereby improving the overall performance for all the tasks. 
In this paper, we exploit a MTL structure to HASA-Net to jointly predict speech quality and intelligibility scores.

\subsection{Multi-head attention mechanism}
\label{ssec:multi-att}
When considering the significance of each speech frame within the input utterances to the target task, the attention mechanism has shown promising effectiveness in making models focus on important information of feature maps. For example, silence frames are less informative and should be given less attention (thus giving lighter weights), whereas frames containing speech are more informative and should be given higher attention (thus giving higher weights) for speech emotion recognition \cite{noh2021multi}. In this study, we use self-attention mechanism to learn different weighted combinations of all time frames in an input sequence. Specifically, an output of the self-attention layer is the weighted sum of the input sequence by calculating every attention weight based on the dependencies between the current frame and other frames within the input sequence. In contrasted to single-head attention, multi-head attention \cite{vaswani2017attention} is adopted to capture information from multiple representation subspaces in order to enhance model's expressiveness. In short, the core idea of incorporating multi-head self-attention in HASA-Net is to emphasize different frame-wise features when predicting speech quality and intelligibility scores. 

\begin{figure}[t!]
\begin{minipage}[b]{0.9\linewidth}
  \centering
  \centerline{\includegraphics[width=7cm]{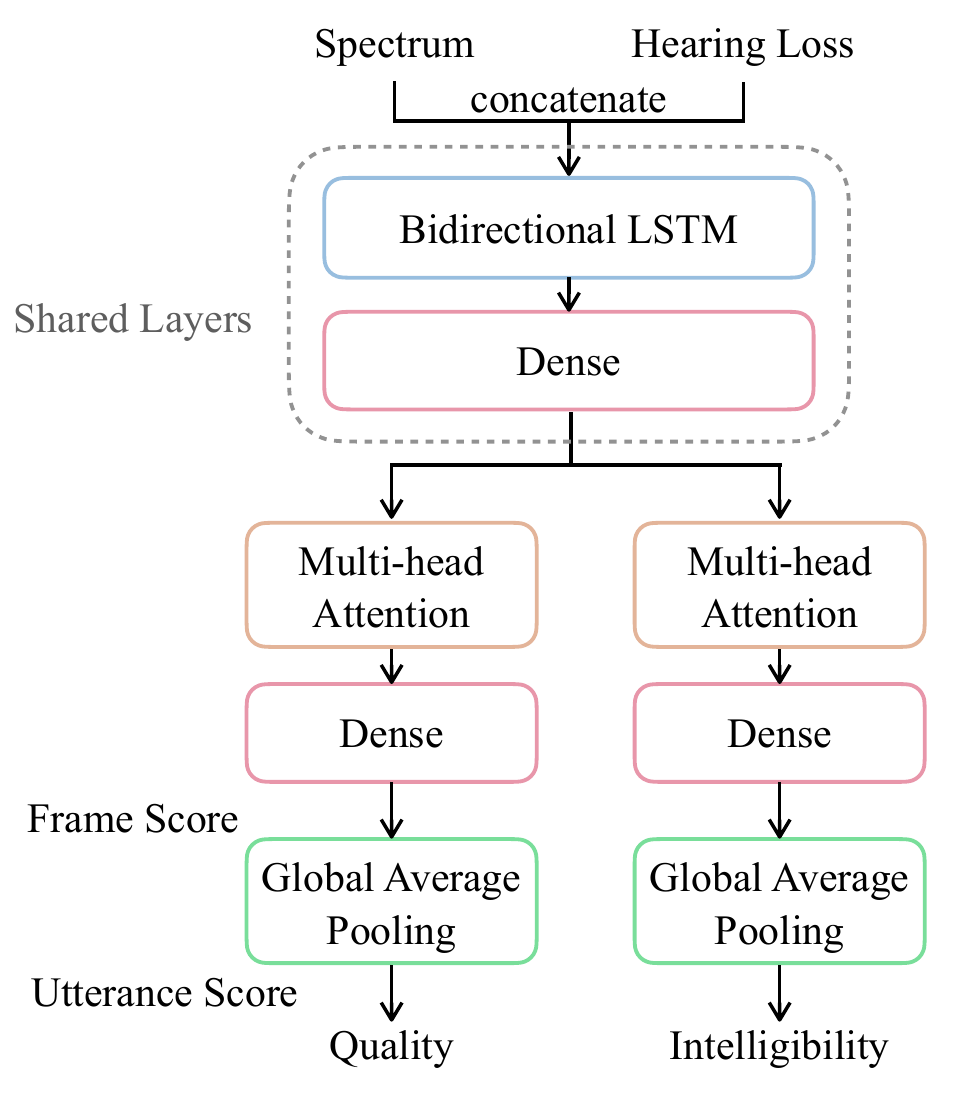}}
\end{minipage}
\caption{Architecture of the HASA-Net model.}
\label{fig:hasanet}
\end{figure}

\subsection{Network architecture}
\label{ssec:network}
Fig.\ref{fig:hasanet} illustrates the overall architecture of HASA-Net. The input of HASA-Net is composed of concatenated Time-frequency (T-F) features and hearing-loss patterns. T-F features are extracted using a 512-point short time Fourier transform (STFT) with a Hamming window size of 512 points and a hop size of 256 points, resulting in a 257-dimension magnitude spectrum. HASA-Net consists of a stack of one bidirectional LSTM with 100 nodes, followed by one dense layer with 128 rectified linear units (ReLU) nodes. By sharing the bidirectional LSTM and dense layers between all tasks, the remaining layers are split into two separate tasks. One task is for quality estimation and the other is for intelligibility prediction. In each task, a multi-head attention mechanism is first applied to aggregate task-specific information based on the shared features. After the multi-head attention, a dense layer with one node activated with a sigmoid function is applied. The output of the dense layer is the predicted score for each frame. Finally, we obtain the final utterance-level score by calculating the global average pooling operation based on frame scores.

\subsection{Objective function}
\label{ssec:objectivefunc}
The loss function for each utterance is the summation of utterance-level loss and averaged frame-wise loss. We formulate the loss for quality estimation as follows:
\begin{displaymath}\label{eq:qualityloss}
\begin{aligned}
L_{Quality} = \frac{1}{N}\sum_{n=1}^{N}[(\hat{Q_n}-Q_n)^2 + \frac{1}{T_n} \sum_{t=1}^{T_n}(\hat{Q_n}-q_{n,t})^2 ]
 \end{aligned}
\end{displaymath}
where $\hat{Q_n}$ and $Q_n$ represent the true and estimated quality scores for the $n$-th utterance, respectively, while $N$ represents the total training utterances and $T_n$ is the number of frames in utterance $n$. Meanwhile, $q_{n,t}$ denotes the estimated frame quality score of the $t$-th frame of utterance $n$. The loss function for the intelligibility estimator is identical to that of quality. Similarly, the loss function for intelligibility estimation is then given as:
\begin{displaymath}\label{eq:intelligibilityloss}
\begin{aligned}
L_{Intelligibility} = \frac{1}{N}\sum_{n=1}^{N}[(\hat{I_n}-I_n)^2 + \frac{1}{T_n} \sum_{t=1}^{T_n}(\hat{I_n}-i_{n,t})^2 ]
\end{aligned}
\end{displaymath}
where $\hat{I_n}$ and $I_n$ denote the true and estimated intelligibility scores for the $n$-th utterance, and $i_{n,t}$ stands for the estimated frame intelligibility score of the $t$-th frame of utterance $n$. 

The overall loss is the sum of the losses for the two tasks:
\begin{displaymath}\label{eq:totalloss}
\begin{aligned}
L_{Total} = \alpha \times L_{Quality}+ \beta \times L_{Intelligibility}
\end{aligned}
\end{displaymath}
where $\alpha$ and $\beta$ represent the weights for the two tasks. According to our internal evaluations, we empirically set $\alpha$ to 1.0 and $\beta$ to 1.5.

\section{Experiments}
\label{sec:experiment}
\subsection{Dataset}
\label{ssec:dataset}
We conducted experiments using the TIMIT database \cite{garofolo1988getting}. All 4620 utterances from the training set of TIMIT were utilized for training. The noisy set was generated by corrupting the utterances with 80 noises combined at seven SNR values:-15, -10, -5, 0, 5, 10 and 15 dB. These noise signals were obtained from the 100 noises dataset \cite{wang2013towards}. In addition, each utterance was corrupted with one noise type at one SNR level so that all the training data was unparalleled. 

For test data, we randomly selected 100 clean speech signals from the test set of the TIMIT database. These selected utterances were then mixed with four unseen noise types (engine, white, street and baby cry) at four SNRs levels:-6,0,6 and 12 dB. In total, the testing data comprises 1600 utterances.

\subsection{Hearing-loss patterns}
\label{ssec:audiograms}
\begin{figure}[t!]
\begin{minipage}[b]{0.32\linewidth}
  \centering
  \centerline{\includegraphics[width=2cm]{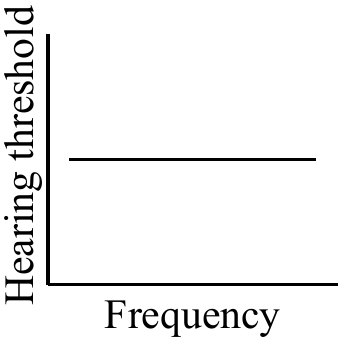}}
  \centerline{(a)}\smallskip
\end{minipage}
\hfill
\begin{minipage}[b]{0.32\linewidth}
  \centering
  \centerline{\includegraphics[width=2cm]{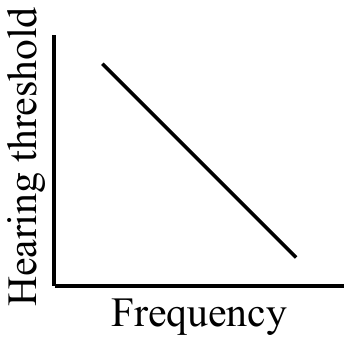}}
  \centerline{(b)}\smallskip
\end{minipage}
\hfill
\begin{minipage}[b]{0.32\linewidth}
  \centering
  \centerline{\includegraphics[width=2cm]{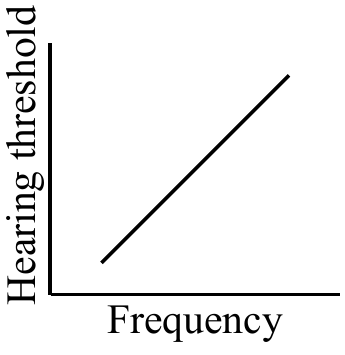}}
  \centerline{(c)}\smallskip
\end{minipage}

\begin{minipage}[b]{0.32\linewidth}
  \centering
  \centerline{\includegraphics[width=2cm]{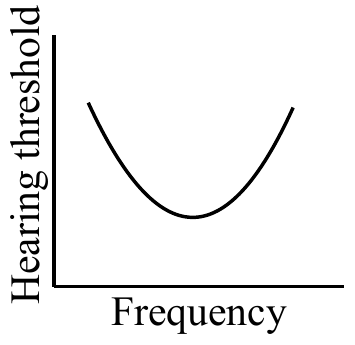}}
  \centerline{(d)}\smallskip
\end{minipage}
\hfill
\begin{minipage}[b]{0.32\linewidth}
  \centering
  \centerline{\includegraphics[width=2cm]{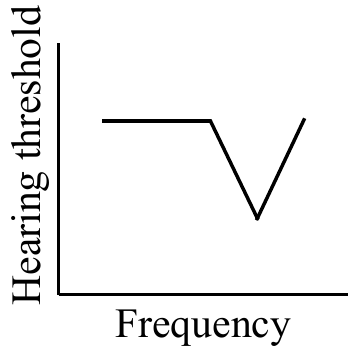}}
  \centerline{(e)}\smallskip
\end{minipage}
\hfill
\begin{minipage}[b]{0.32\linewidth}
  \centering
  \centerline{\includegraphics[width=2cm]{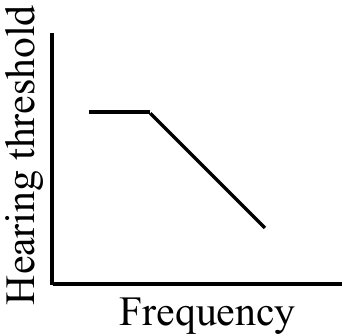}}
  \centerline{(f)}\smallskip
\end{minipage}
\caption{Hearing loss configurations. (a) Flat (b) Sloping (c) Rising (d) Cookie-bite (e) Noise-notched (f) High-frequency.}
\label{fig:fig_hl}
\end{figure}

Hearing loss is detected through an audiogram, which is a graphical display showing the degrees of hearing loss at different frequency regions, with the y-axis representing the hearing threshold and the x-axis representing the frequency. A threshold at any frequency above 20 dB is considered as a hearing loss. The hearing loss patterns are selected at frequencies of 250, 500, 1000, 2000, 4000, and 6000 Hz. Each audiogram is assigned to one of six categories flat, sloping, rising, cookie-bite, noise-notched and high-frequency \cite{alshuaib2015classification}. Different configurations of hearing loss are shown in Fig. \ref{fig:fig_hl}. 

In the flat category (Fig. \ref{fig:fig_hl}(a)), there are no change in thresholds across frequencies. Audiograms with sloping configurations have increased thresholds from low to high frequencies (Fig. \ref{fig:fig_hl}(b)). The rising category with thresholds decrease as frequencies increased (Fig. \ref{fig:fig_hl}(c)). A cookie-bite audiogram has greatest thresholds in the mid frequencies between 500 and 4000 Hz (Fig. \ref{fig:fig_hl}(d)). A noise-notch configuration reaches a maximum threshold in frequency between 3000 and 6000 Hz (Fig. \ref{fig:fig_hl}(e)). High-frequency category has hearing loss occurs at frequencies between 2000 and 8000 Hz, while frequencies below 2000 Hz are not affected (Fig. \ref{fig:fig_hl}(f)). 

In total, we have collected 42 hearing loss patterns, where each hearing loss category contained 7 patterns. Every category was further divided into two groups, the larger group with 5 patterns and the smaller one with the remaining 2 patterns.  

\subsection{Experimental setup}
\label{ssec:setup}
As mentioned earlier, HASA-Net takes the magnitude spectrum and the hearing-loss patterns as its inputs and outputs the corresponding quality and intelligibility scores. Given a training utterance, two hearing-loss patterns from the mentioned larger group were selected to form the corresponding training set, which resulted in 4158 $\times$ 2 training utterances in every category.

To evaluate HASA-Net's generalization capability, we considered two scenarios: seen and unseen test sets. The seen test sets utilized the hearing-loss patterns selected from the larger groups and were already presented in the training, whereas the unseen test set contained patterns from the smaller groups that were not used in the training. Each category included two hearing-loss patterns for seen and unseen test sets, and both of them included the same testing utterances. Therefore, there were 19,200 utterances in each of the seen and unseen test sets.

The corresponding ground-truth values for quality and intelligibility of HASA-Net were those calculated by HASQI and HASPI, respectively. HASQI and HASPI are based on an auditory model that incorporates changes due to hearing loss, and then develop mathematical models that match the corresponding quality and intelligibility scores. Both of their metrics are scalars between 0 and 1, where a higher score of HASQI and HASPI represents better speech quality and intelligibility. The stimuli were amplified using the National Acoustics Laboratories revised (NAL-R) \cite{byrne1986national} linear fitting prescriptive formula based on individual hearing-loss patterns.  

We use the RMSprop \cite{tieleman2012lecture} optimizer with a learning rate of 0.001 and early stopping technique to train HASA-Net. To evaluate the performance, various criteria including mean square error (MSE), linear correlation coefficient (LCC), and Spearman’s rank correlation coefficient (SRCC) are selected. In the follows, we first show the detailed assessment results of HASA-Net.

\subsection{Detailed assessment results }
\label{ssec:result}

\begin{figure}[ht!]
\begin{minipage}[b]{0.48\linewidth}
  \centering
  \centerline{\includegraphics[width=4.25cm]{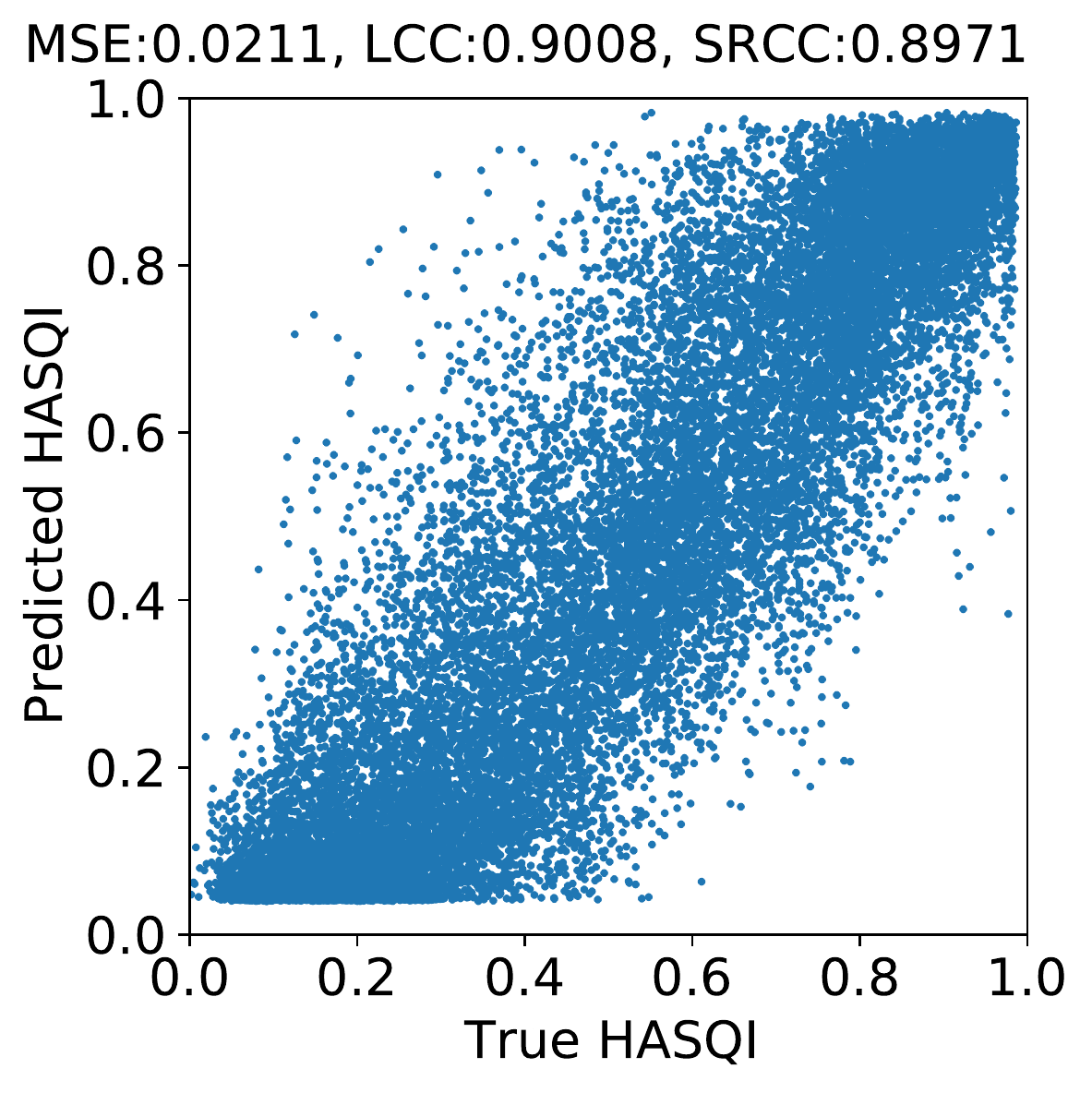}}
  \centerline{(a)}\smallskip
\end{minipage}
\hfill
\begin{minipage}[b]{0.48\linewidth}
  \centering
  \centerline{\includegraphics[width=4.25cm]{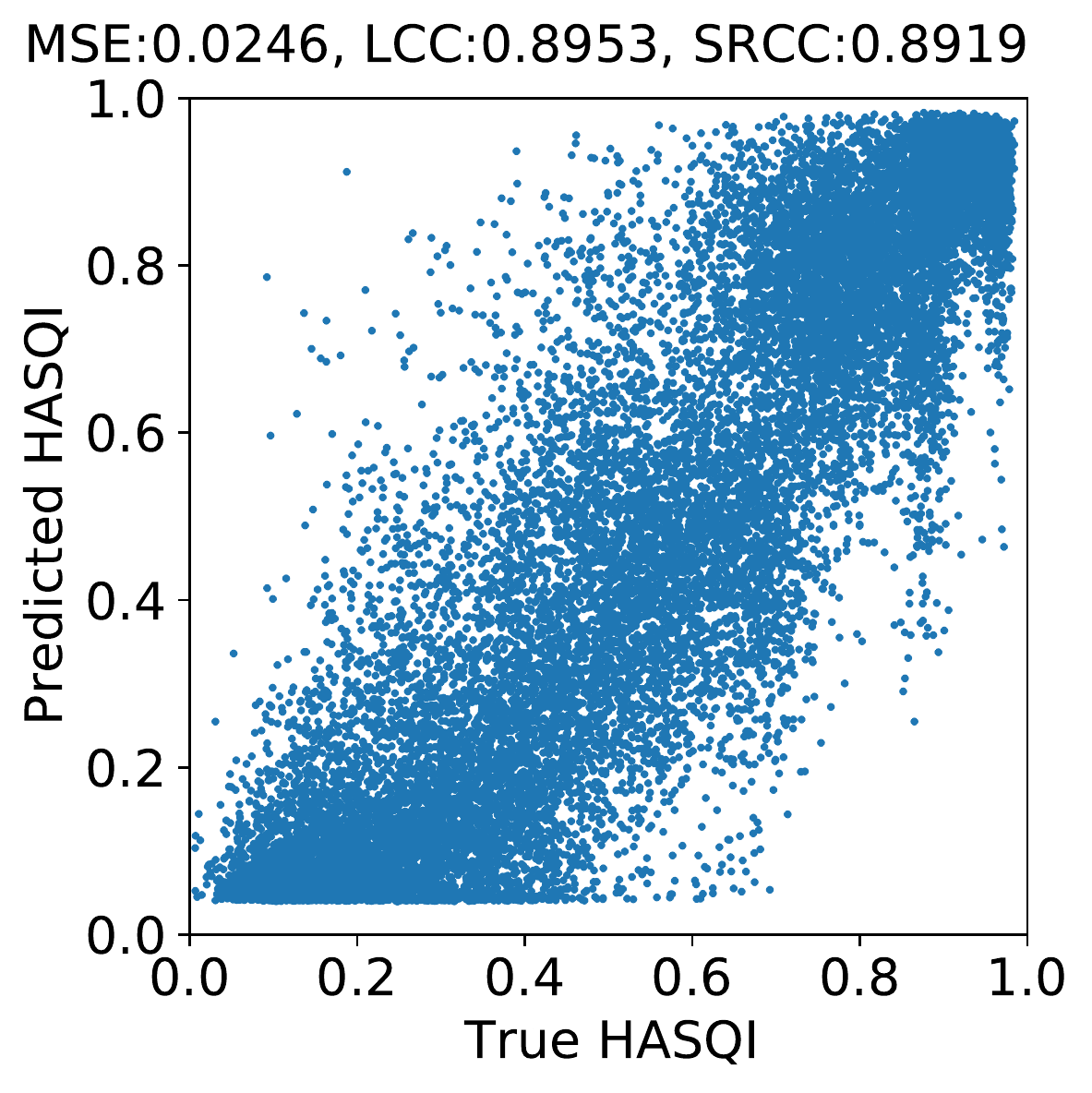}}
  \centerline{(b)}\smallskip
\end{minipage}
\caption{Scatter plots for quality assessment by HASA-Net. (a) seen test set (b) unseen test set.}
\label{fig:fig_hasqi}
\end{figure}

\begin{figure}[ht!]
\begin{minipage}[b]{0.48\linewidth}
  \centering
  \centerline{\includegraphics[width=4.25cm]{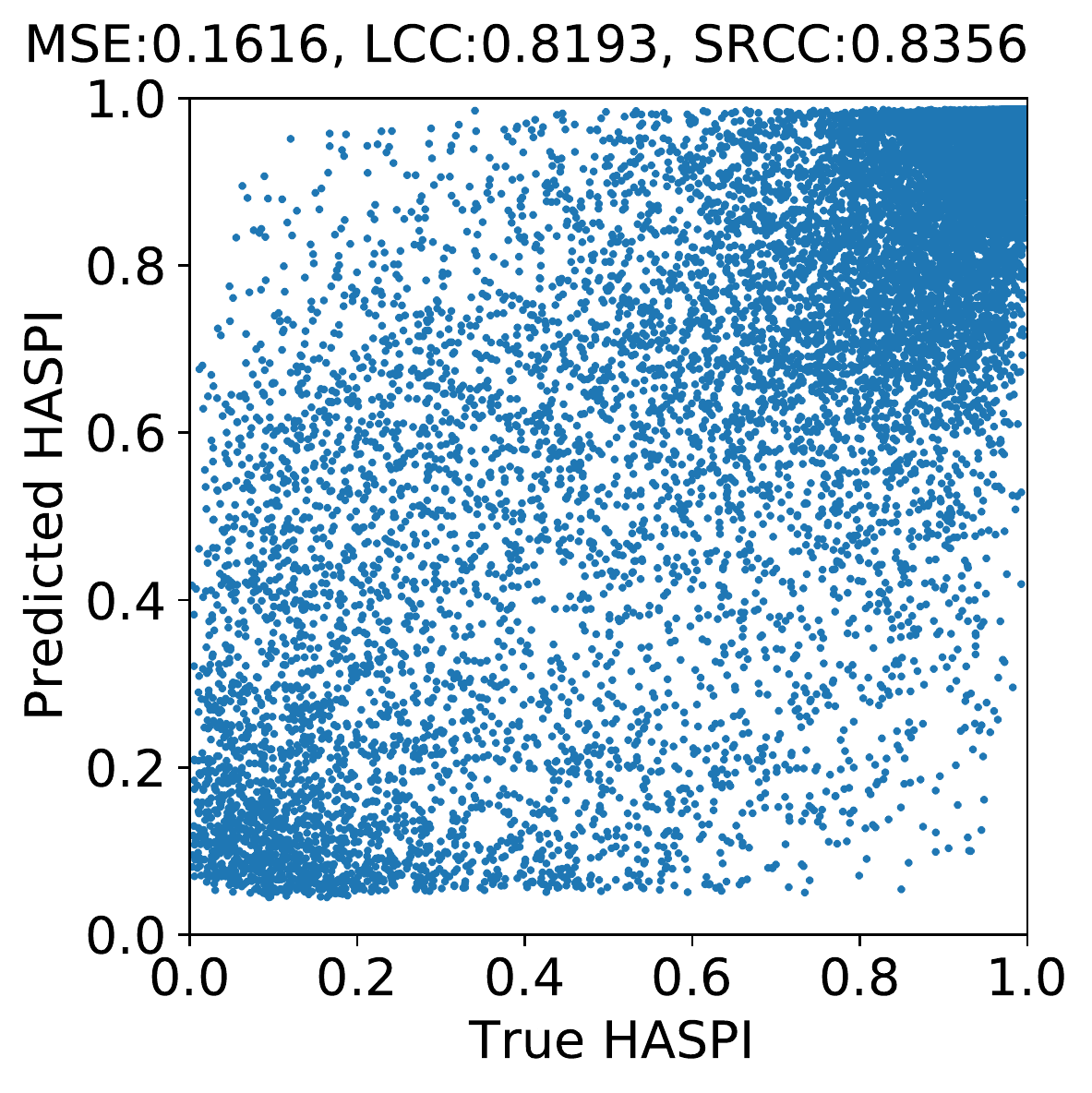}}
  \centerline{(a)}\smallskip
\end{minipage}
\hfill
\begin{minipage}[b]{0.48\linewidth}
  \centering
  \centerline{\includegraphics[width=4.25cm]{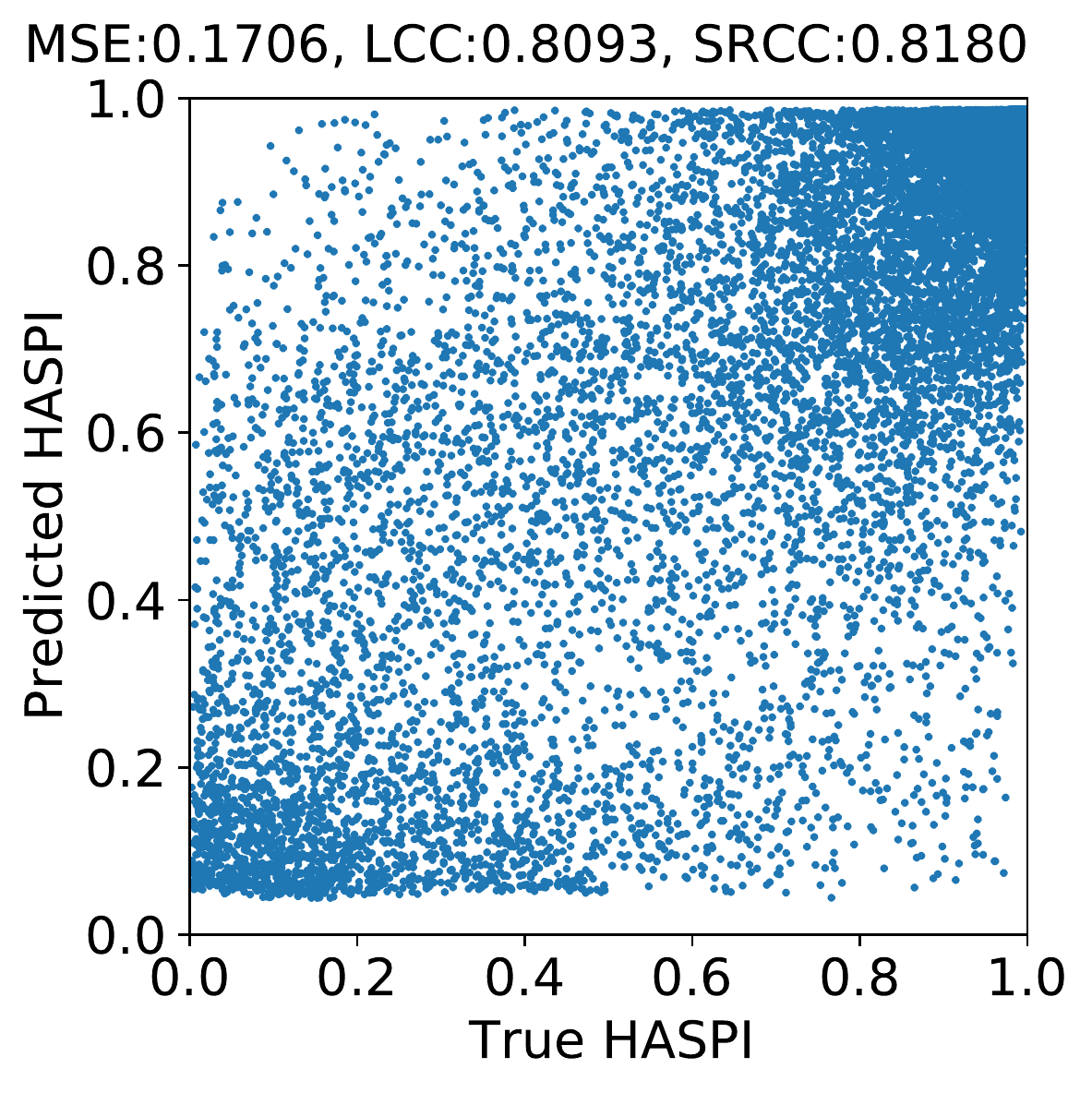}}
  \centerline{(b)}\smallskip
\end{minipage}
\caption{Scatter plots for intelligibility assessment by HASA-Net. (a) seen test set (b) unseen test set.}
\label{fig:fig_haspi}
\end{figure}

We present the detailed assessment results of HASA-Net for quality and intelligibility on the seen and unseen test sets. Figs. \ref{fig:fig_hasqi} and \ref{fig:fig_haspi} show the scatter plots with the corresponding quality and intelligibility metrics. Each figure shows results from the seen and unseen test sets. The higher the LCC and SRCC, the more accurately HASA-Net predicts the corresponding ground truths, indicating higher prediction accuracies. On average, HASA-Net achieves LCC of 0.9008 and SRCC of 0.8971 on the seen test set for the quality estimation, compared to 0.8953 and 0.8919 on the unseen test set. For the intelligibility prediction, \{LCC, SRCC\} are \{0.8193, 0.8356\} and \{0.8093, 0.818\} on the seen and unseen test sets, respectively. In our experiments, HASA-Net shows a strong correlation between the prediction scores and the ground truths with LCC and SRCC around 0.89 to 0.9 for the quality estimation, and a decrease that lies between 0.8 to 0.84 for the intelligibility estimation. 

\begin{figure}[ht!]
\begin{minipage}[b]{0.95\linewidth}
  \centering
  \centerline{\includegraphics[width=6.5cm]{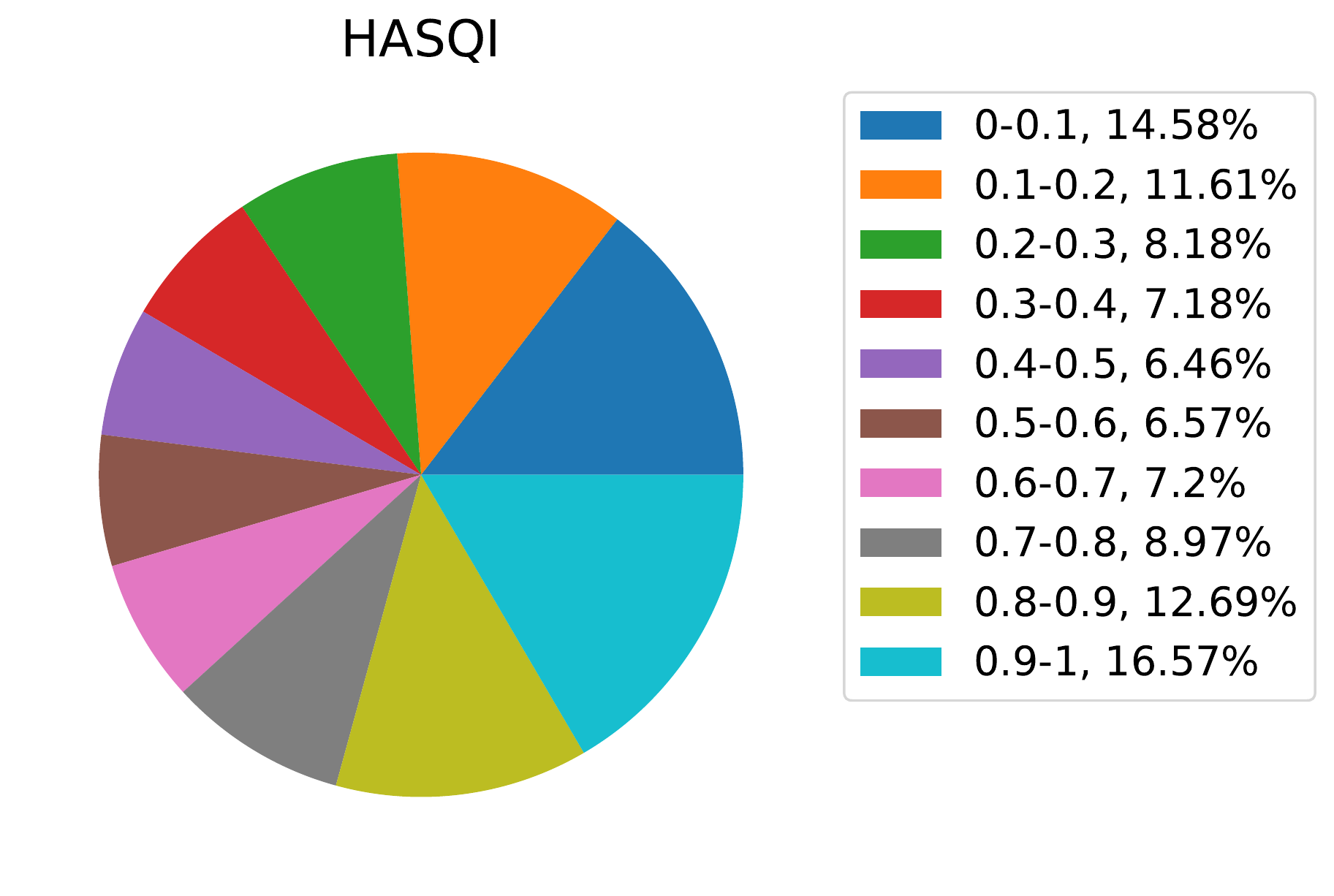}}
  \centerline{(a)}\smallskip
\end{minipage}
\begin{minipage}[b]{0.95\linewidth}
  \centering
  \centerline{\includegraphics[width=6.5cm]{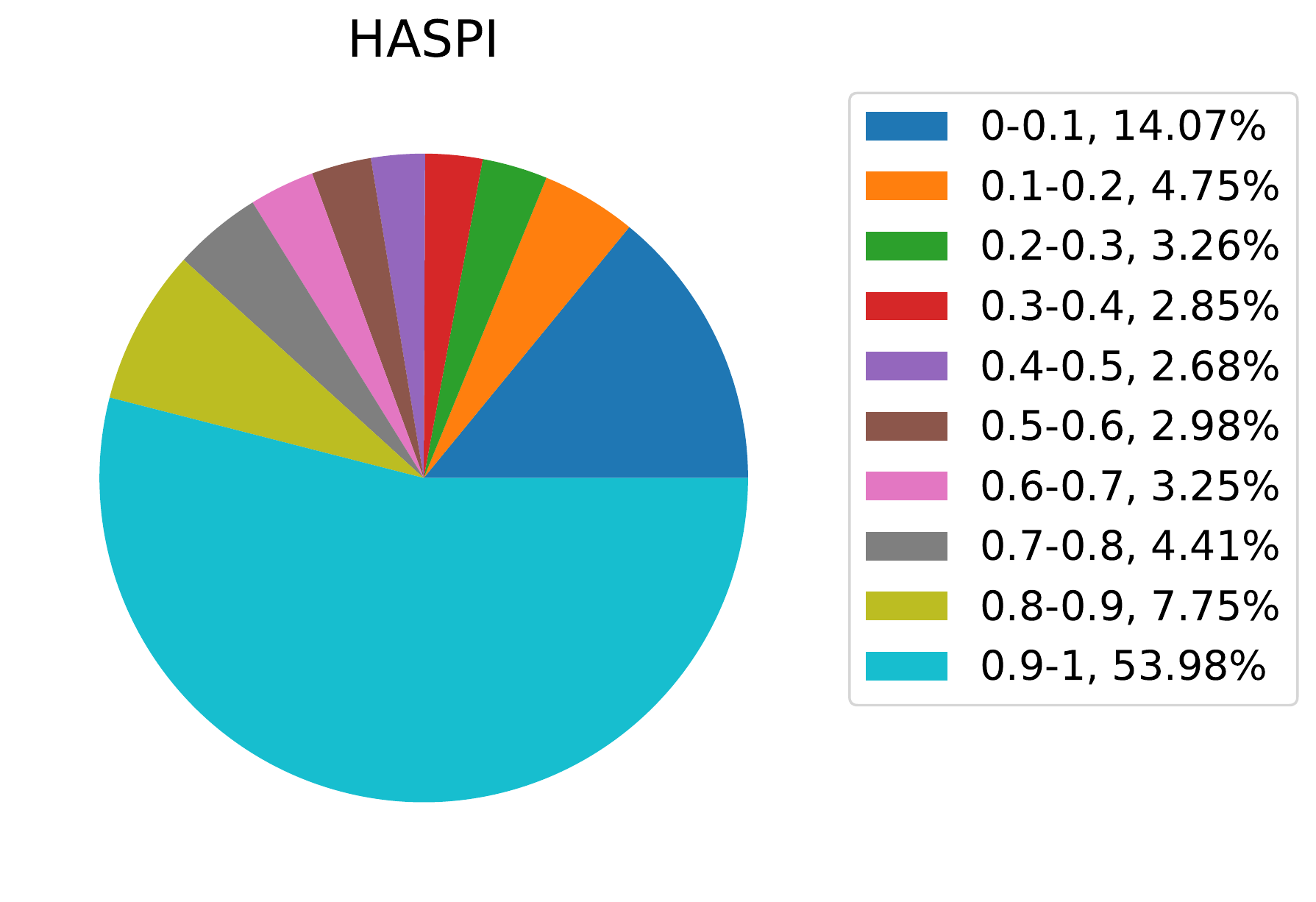}}
  \centerline{(b)}\smallskip
\end{minipage}
\caption{Pie charts depicting the percentage allocation of quality and intelligibility scores according to training set. (a) HASQI (b) HASPI.}
\label{fig:fig_cdf}
\end{figure}

The above results show that intelligibility is more difficult to evaluate than quality, which may be caused by differences in the sensitivity of HASQI and HASPI. The authors in \cite{kates2018using} pointed out that HASQI is much more sensitive to noise and show that at 10 dB SNR level, HASPI is 0.99 which is almost close to 1, whereas HASQI is only about 0.3. This is consistent with what we have found in our training set where HASPI scores are much more likely to be close to 1 than HASQI scores. Fig. \ref{fig:fig_cdf} illustrate the percentage allocation of quality and intelligibility scores in the training set. Provided that the same spectrum and hearing-loss patterns are used, HASQI scores are distributed evenly from 0 to 1, while around 14\% and 54\% of HASPI scores fall in the range of 0 to 0.1 and 0.9 to 1. This uneven HASPI score distribution of training set brings difficulty to training process. Although HASA-Net achieves slightly lower prediction performance of intelligibility compared to that of quality, it is still promising that HASA-Net has a great chance to estimate values close to the true intelligibility with average LCC and SRCC of 0.8143 and 0.8268. 

Table \ref{table:table_hasqi} and \ref{table:table_haspi} provide further details of quality and intelligibility prediction performance based on different types of hearing loss. As mentioned above, HASA-Net's prediction of intelligibility is not as good as its prediction of quality as measured by LCC and SRCC on the seen and unseen test sets. However, in the worse case, HASA-Net still achieves \{LCC, SRCC\} around \{0.7392, 0.7259\} for the unseen rising hearing-loss patterns, which indicates that the estimated scores still follows the trend of the true scores. In addition, similar tendencies can be found in the quality and intelligibility estimation. We observe that sloping configuration shows higher LCCs and SRCCs, while lower LCCs and SRCCs are found in rising and cookie-bite configurations.  

\begin{table}[t]
    \caption{Performance of quality prediction of different types of hearing loss on seen and unseen test sets.}
    \centering
    \resizebox{\columnwidth}{!}{%
    \begin{tabular}{lcccccc}
    \toprule
    &\multicolumn{3}{c}{Seen}&\multicolumn{3}{c}{Unseen}\\
    \cmidrule(r){2-4}\cmidrule(r){5-7} 
    Configuration & MSE & LCC & SRCC & MSE & LCC & SRCC \\
    \midrule
    Flat & 0.0221 & 0.8975 & 0.8923 & 0.026 & 0.89 & 0.8806\\
    Sloping & 0.0232 & 0.9177 & 0.8959 & 0.0199 & 0.9103 & 0.894 \\
    Rising  & 0.0213 & 0.8805 & 0.8641 & 0.0242 & 0.8834 & 0.8719\\
    Cookie-bite & 0.0234 & 0.8884 & 0.8836 & 0.0342 & 0.8979 & 0.894\\
    Noise-notched & 0.0177 & 0.902 & 0.9019 & 0.0247 & 0.8716 & 0.88\\
    High-frequency &0.0192 & 0.8982 & 0.8862 & 0.0186 & 0.9094 & 0.8907\\
    \bottomrule
    \end{tabular}
    }
    \label{table:table_hasqi}
\end{table}

\begin{table}[htb!]
    \caption{Performance of intelligibility prediction of different types of hearing loss on seen and unseen test sets.}
    \centering
    \resizebox{\columnwidth}{!}{%
    \begin{tabular}{lcccccc}
    \toprule
    &\multicolumn{3}{c}{Seen}&\multicolumn{3}{c}{Unseen}\\
    \cmidrule(r){2-4}\cmidrule(r){5-7} 
    Configuration & MSE & LCC & SRCC & MSE & LCC & SRCC \\
    \midrule
    Flat & 0.0226 & 0.8034 & 0.885 & 0.0313 & 0.8103 & 0.8298\\
    Sloping & 0.0312 & 0.8324 & 0.7823 & 0.0224 & 0.8622 & 0.8167 \\
    Rising  & 0.0474 & 0.7957 & 0.7983 & 0.0516 & 0.7392 & 0.7259\\
    Cookie-bite & 0.0273 & 0.7338 & 0.8259 & 0.0399 & 0.7829 & 0.8129\\
    Noise-notched & 0.0189 & 0.855 & 0.8238 & 0.016 & 0.8205 & 0.8471\\
    High-frequency & 0.0258 & 0.8533 & 0.8004 & 0.0296 & 0.8451 & 0.7746\\
    \bottomrule
    \end{tabular}
    }
    \label{table:table_haspi}
\end{table}

Finally, HASA-Net achieves comparable results in both quality and intelligibility on the seen and unseen test sets across different types of hearing loss. Not surprisingly, the seen test set outperforms the unseen for most types of hearing loss. However, when considering these challenging test conditions where the hearing-loss patterns are totally unseen, the performance of HASA-Net continues to be quite strong, proving the robustness and generalization capability of HASA-Net. In summary, our detailed analysis confirms the effectiveness of HASA-Net. 

\subsection{Single-task versus HASA-Net}
\label{ssec:stnmtl}
In this section, we analyze the role of MTL. A single-task model adopts the same architecture of HASA-Net, but outputs only one objective prediction. Table \ref{table:table_hasqi_mtk} and \ref{table:table_haspi_mtk} show the comparison results between the single-task model and HASA-Net. From the tables, we can see that HASA-Net produces better results compared to the single-task model, showing the benefit of using MTL. For quality prediction, notable improvement from the single-task model to HASA-Net can be noted. For instance, HASA-Net produces \{LCC, SRCC\} improvements of \{0.0314, 0.0269\} and \{0.0303, 0.0253\} for the seen and unseen test sets, respectively, as compared to the single-task model. On the other hand, intelligibility prediction shows relatively less improvement. Specifically, HASA-Net achieves \{LCC, SRCC\} improvements of \{0.0083, 0.0258\} and \{0.0122, 0.0575\} in comparison to single-task model for the seen and unseen test sets. The resutls show that MTL assists HASA-Net to learn more robust and universal feature representations across different tasks. Instead of overfitting to a specific task, MTL scheme allows for learning task-specific information from the shared feature representations to further boost the performance. In contrast to the single-task model, an addition advantage of HASA-Net is its ability to predict quality and intelligibility scores simultaneously. For applications with limited computation and storage resources, HASA-Net is definitely a better choice than multiple single-task models. 

\begin{table}[t]
    \caption{Performance of quality prediction between single-task and HASA-Net on seen and unseen test sets.}
    \centering
    \resizebox{\columnwidth}{!}{%
    \begin{tabular}{lcccccc}
    \toprule
    &\multicolumn{3}{c}{Seen}&\multicolumn{3}{c}{Unseen}\\
    \cmidrule(r){2-4}\cmidrule(r){5-7} 
     & MSE & LCC & SRCC & MSE & LCC & SRCC \\
    \midrule
    Single-task & 0.0238 & 0.8694 & 0.8702 & 0.0267 & 0.865 & 0.8666\\
    HASA-Net & \textbf{0.0211} & \textbf{0.9008} & \textbf{0.8971} & \textbf{0.0246} & \textbf{0.8953} & \textbf{0.8919} \\
    \bottomrule
    \end{tabular}
    }
    \label{table:table_hasqi_mtk}
\end{table}

\begin{table}[ht!]
    \caption{Performance of intelligibility prediction between single-task and HASA-Net on seen and unseen test sets.}
    \centering
    \resizebox{\columnwidth}{!}{%
    \begin{tabular}{lcccccc}
    \toprule
    &\multicolumn{3}{c}{Seen}&\multicolumn{3}{c}{Unseen}\\
    \cmidrule(r){2-4}\cmidrule(r){5-7} 
     & MSE & LCC & SRCC & MSE & LCC & SRCC \\
    \midrule
    Single-task & 0.0312 & 0.8110 & 0.8098 & 0.0342 & 0.7971 & 0.7605\\
    HASA-Net & \textbf{0.0289} & \textbf{0.8193} & \textbf{0.8356} & \textbf{0.0318} & \textbf{0.8093} & \textbf{0.8180} \\
    \bottomrule
    \end{tabular}
    }
    \label{table:table_haspi_mtk}
\end{table}

\section{Conclusion}
\label{sec:conclusion}
In this study, we have proposed a DNN-based speech assessment model for listeners using hearing aids, which we call HASA-Net. To the best of our knowledge, this is the first end-to-end, non-intrusive unified model predicting hearing aid speech quality and intelligibility simultaneously. Our experimental results show that HASA-Net's quality and intelligibility estimates are highly correlated with HASQI and HASPI scores. We also demonstrated the superior performance of HASA-Net as compared to the single-task model. The end-to-end framework gave HASA-Net the capability to directly combine with DNN-based speech signal processing models for joint optimization. HASA-Net provides a significant building block towards tackling various complex challenges for the hearing-impaired, such as speech enhancement, dereverberation, and separation. Future work includes listening tests with hearing-impaired listeners to investigate correlation of HASA-Net and human ratings.

\bibliographystyle{IEEEbib}
\bibliography{refs}
\end{document}